\documentclass[journal]{IEEEtran}
\usepackage{cite}
\ifCLASSINFOpdf
   \usepackage[pdftex]{graphicx}
  \graphicspath{{/home/jake/Documents/LaTeX_Files/}{./TeX_Figs/}}
\else
   \usepackage[dvips]{graphicx}
  \graphicspath{{/home/jake/Documents/LaTeX_Files/}{./TeX_Figs/}}
\fi
\usepackage{amssymb}

\usepackage[cmex10]{amsmath}
\usepackage{array}
\usepackage{mdwmath}
\usepackage{mdwtab}
\usepackage{fixltx2e}
\newcommand{\mbb}{\mathbb}

\newcommand{\mc}{\mathcal}

\newcommand{\p}{\prime}

\newcommand{\subeq}{\subseteq}

\newcommand{\mcp}{\mathcal{P}}

\newcommand{\tbf}{\textbf}

\newtheorem{theorem}{Theorem}[section]

\newtheorem{proposition}[theorem]{Proposition}

\begin{document}
\title{Quantum LDPC Codes Constructed from Point-Line Subsets of the Finite Projective Plane}

\author{Jacob~Farinholt
\thanks{Manuscript received Month Day, Year; revised Month Day, Year.}%
\thanks{This work was supported by the Naval Surface Warfare Center's In-house Laboratory Independent Research (ILIR) and Academic Fellowship (AFP) programs.}%
\thanks{J. Farinholt is with the Electromagnetic and Sensor Systems Department, Naval Surface Warfare Center, Dahlgren Division, Dahlgren, VA 22448 (e-mail:Jacob.Farinholt@navy.mil)}}

\markboth{Journal Name,~Vol.~\#, No.~\#, June~2012}%
{Quantum LDPC Codes Constructed from Point-Line Subsets of the Finite Projective Plane}
%{Shell \MakeLowercase{\textit{et al.}}: Bare Demo of IEEEtran.cls for Journals}



\maketitle

\begin{abstract}
%\boldmath
Due to their fast decoding algorithms, quantum generalizations of low-density parity check, or LDPC, codes have been investigated as a solution to the problem of decoherence in fragile quantum states.  However, the additional twisted inner product requirements of quantum stabilizer codes force four-cycles and eliminate the possibility of randomly generated quantum LDPC codes.  Moreover, the classes of quantum LDPC codes discovered thus far generally have unknown or small minimum distance, or a fixed rate.  This paper presents several new classes of quantum LDPC codes constructed from finite projective planes.  These codes have rates that increase with the block length $n$ and minimum weights proportional to $n^{1/2}$.
\end{abstract}


% \fi
\IEEEpeerreviewmaketitle

\nocite{JFarin12}

\section{Introduction}
\IEEEPARstart{C}{lassical} low-density parity check codes, or LDPC codes, were first discovered by Gallager \cite{Gal62} in 1960. Later, it was shown that bipartite graphs, called \emph{Tanner graphs}, could be used to describe the codes and their actions under iterative belief propagation decoding algorithms. In 2001, Kou \emph{et al.} \cite{KLF01} showed that by using finite geometries, many classes of these codes could be easily generated to have known parameters. Recently, Droms \emph{et al.} \cite{DrMel06} and Castleberry \emph{et al.} \cite{CHMel10} showed that LDPC codes constructed from point-line subsets of finite projective planes could often out-perform those constructed in \cite{KLF01}. Classical LDPC codes are some of the best known, with rates asymptotically approaching the Shannon limit \cite{Shan48}.  We say a code is an LDPC code if its parity check matrix is sparse, and its corresponding Tanner graph has no four-cycles (i.e., any pair of rows in the parity check have no more than one ``1'' in common position). In particular, the rows of the parity check matrix need not all be linearly independent.

The first quantum error correcting codes were discovered by Shor \cite{Shor95}, Calderbank \cite{CalShor96}, and separately by Steane \cite{Steane, Stea96}.  Using the stabilizer formalism introduced by Gottesman \cite{Gott96}, it was shown that these corresponding \emph{stabilizer codes} could be described using classical parity check matrices with an added twisted inner product requirement \cite{CalRainsShorSloane}. This twisted inner product has the unfortunate consequence of forcing four-cycles on a corresponding Tanner graph.  Nevertheless, sparse-graph stabilizer codes with minimal four-cycles have been suggested as quantum generalizations of classical LDPC codes. The first such quantum LDPC, or QLDPC, codes were suggested by Postol in 2001 \cite{Pos01}, and many examples were constructed by MacKay \emph{et al.} \cite{MkMf04}. Since then, many other classes of QLDPC codes have been constructed (see \cite{TilZem09, Aly09} and references therin). However, many of these codes have unknown or small minimum distance. Tillich and Zemor \cite{TilZem09} constructed fixed-rate QLDPC codes with minimum distances that increase as the square root of the code length. Aly \cite{Aly09} used finite geometric techniques to create quantum generalizations of many of the classical LDPC codes constructed in \cite{KLF01}, and suggested that similar codes could be constructed through the use of projective geometries. In this paper, projective geometries, in particular the projective plane of order $2^s$ and many of its subsets, are used to construct QLDPC codes with minimum distances proportional to the square root of the code length, and whose rates also increase with the code length, providing possibly the best-known rate-increasing QLDPC codes in the current literature.

This paper is organized as follows. In Section \ref{CSS} we introduce quantum stabilizer codes, classical constructions, and the CSS formalism. In Section \ref{PG(2,q)} we introduce finite projective planes and subsets of the plane with respect to regular hyperovals. Section \ref{ClassicalLDPC} describes methods of constructing classical self-orthogonal sparse-graph codes from the subsets described in Section \ref{PG(2,q)}. Lastly, Section \ref{QLDPC} contains our main results about new classes of Quantum LDPC codes and their corresponding parameters.

%%%%%%%%%%%%%%%%%%%%%%%%%%%%%%%%%%%%%%%%%%%%%%%%%%%%%%%%%%%
%%%%%%%%%%%%%%%%%%%%%%%%%%%%%%%%%%%%%%%%%%%%%%%%%%%%%%%%%%%
\section{Quantum Stabilizer Codes and CSS Codes}\label{CSS}

The Pauli operators are given by 
\begin{align}\label{Eq:Pauli}
I &= \begin{pmatrix} 1 & 0 \\ 0 & 1 \end{pmatrix}, \ &X = \begin{pmatrix} 0&1\\1&0 \end{pmatrix}, \notag\\
Y &= \begin{pmatrix}0&-i\\i&0\end{pmatrix}, \ &Z = \begin{pmatrix}1&0\\0&-1\end{pmatrix}.
\end{align}
These operators, along with the scalars $i^k$ for $k \in \mathbb{Z}_4$, form the single-qubit \emph{Pauli group} $\mc{P}$. Observe that $X^2 = Y^2 = Z^2 = I$, and $Y = iXZ$. A pure state quantum bit, or \emph{qubit} is a norm-one element of a two-dimensional complex Hilbert space, $\mc{H}_2$.  It is well-known \cite{Steane} that correcting errors induced by the Pauli operators is sufficient to correct arbitrary errors on single qubits.

In the case of multi-qubit states, that is, norm-one elements in $\mc{H}_2^{\otimes n} = \mc{H}_2 \otimes \mc{H}_2 \dots \otimes \mc{H}_2$, the $n$-fold tensor product of $\mc{H}_2$, the corresponding Pauli operators are $n$-fold tensor products of the operators given in \eqref{Eq:Pauli}, with corresponding Pauli group $\mc{P}_n$ on $n$ qubits obtained by including the scalars $i^k$ for $k \in \mathbb{Z}_4$.

An arbitrary element $P \in \mc{P}_n$ is represented as
\begin{equation}\label{PauliElement}
  P = i^kX(a)Z(b),
\end{equation}
for $k \in \mbb{Z}_4$, and $a = (a_1, a_2, \dots, a_n)$ and $b = (b_1, b_2, \dots, b_n)$ are both length $n$ binary vectors, with a 1 in position $i$ of $a$ (resp. $b$) precisely when there is an $X$ (resp. $Z$) operator acting on qubit $i$.  For example, $X(1001)Z(0101)$ should be interpreted as $X\otimes Z\otimes I \otimes Y$, with the $Y$ at the end since it is a product of $X$ and $Z$. It has been shown \cite{CalRainsShorSloane} that two elements $i^kX(a)Z(b)$ and $i^{k^\p}X(a^\p)Z(b^\p)$ of $\mc{P}_n$ commute if and only if
\begin{equation}\label{Eq:TIP}
  (a|b) * (a^\p|b^\p) := a \cdot b^\p + b \cdot a^\p = 0 \pmod 2,
\end{equation}
where $\cdot$ is the standard inner (dot) product. We call $*$ the \emph{twisted inner product}.

Observe that the center $C(\mc{P}_n)$ of $\mc{P}_n$ is given by $C(\mcp_n) = \{i^kI \ | \ k \in \mbb{Z}_4 \}$. Since these elements are effectively the global phase actions in $\mcp_n$, we can reduce ourselves to considering the quotient group $\mcp_n/C(\mcp_n)$, the elements of which are equivalence classes of the form $\{i^kX(a)Z(b) \ | \ k \in \mbb{Z}_4 \}$, for fixed $a$ and $b$; and we label each equivalence class with its scalar-free element (e.g. $X(a)Z(b) \equiv \{i^kX(a)Z(b) \ | \ k \in \mbb{Z}_4 \}$). Moreover, as described in \cite{CalRainsShorSloane}, the quotient group $\mcp_n/C(\mcp_n)$ is isomorphic to a $2n$-dimensional binary vector space $V_{2n}$ via the map
\begin{equation}
  X(a)Z(b) \mapsto (a|b),
\end{equation}
with the commutativity relationship of elements in $\mcp_n$ preserved by imposing the twisted inner product \eqref{Eq:TIP} on the vector space $V_{2n}$.

%%%%%%%%%%%%%%%%%%%%%%%%%%%%%%%%%%%%%%%%%%%%%%%%%%%%%%%%%%%%
\subsection{Stabilizer Codes}\label{Subsec:StabCodes}
A \emph{stabilizer group} $\mc{S} \subeq \mcp_n$ is a commutative subgroup of $\mcp_n$ that does not contain $-I$.  By not containing $-I$, distinct elements of $\mc{S}$ are mapped to distinct equivalence classes in $\mcp_n/C(\mcp_n)$.  Thus, without loss of generality, a stabilizer group can be classically represented as a collection of vectors in $V_{2n}$ with the property that every pair are orthogonal with respect to the twisted inner product \eqref{Eq:TIP}.

A \emph{stabilizer code} $\mc{C}(\mc{S})$ for stabilizer group $\mc{S} \subeq \mcp_n$ is defined as the simultaneous $+1$ eigenspace in $\mc{H}_2^{\otimes n}$ of each element in $\mc{S}$.  An error is detected if it anticommutes with any element of the stabilizer group, thereby producing a nonzero syndrome. Thus, we can classically define a stabilizer parity check matrix $H(\mc{C})$ for the stabilizer code $\mc{C}(\mc{S})$ by letting the rows of $H(\mc{C})$ be the vectors corresponding to the generators of $\mc{S}$.

Conversely, a binary length-$2n$ matrix $[A|B]$ is a stabilizer parity check matrix for some stabilizer code if and only if every pair of rows in $[A|B]$ are orthogonal with respect to \eqref{Eq:TIP}, or equivalently, if and only if
\begin{equation}
  AB^t + BA^t = 0 \pmod 2.
\end{equation} 

%%%%%%%%%%%%%%%%%%%%%%%%%%%%%%%%%%%%%%%%%%%%%%%%%%%%%%%%%%%%
\subsection{CSS Codes}\label{Subsec:CSS}
A large class of stabilizer codes of particular interest are ones that are constructed from classical error correcting codes.  In particular, if $C_1$ and $C_2 \subeq C_1^\bot$ are classical codes with parity check matrices $H(C_1)$ and $H(C_2)$, respectively, then
\begin{equation}\label{Eq:AsymCSS}
  \tbf{H}(\mc{C}) = \begin{pmatrix} H(C_1) & 0\\ 0 & H(C_2)\end{pmatrix}
\end{equation}
is a stabilizer parity check for a stabilizer code $\mc{C}$.  Note that \eqref{Eq:TIP} is satisfied by the fact that $H(C_1)$ is orthogonal to $H(C_2)$ by construction. In particular, if $C_1$ and $C_2$ are classical $[n,k_1,d_1]$ and $[n,k_2,d_2]$ codes, respectively, then the stabilizer code $\mc{C}$ encodes $K = k_1+k_2 - n$ qubits into $n$ qubits, and corrects $(D-1)/2$ and fewer arbitrary qubit errors, where $D = \min(d_1, d_2)$ \cite{CalRainsShorSloane}.  We call such a code an $[[n, K, D]]$ CSS code, where CSS are the initials of the discoverers of such codes \cite{CalShor96, Stea96}. Note that it is standard practice to place the parameters of a quantum code inside double brackets, as opposed to single brackets in the classical case.  A particularly nice scenario is one in which a classical $[n, k, d]$ code $C$, with parity check matrix $H(C)$, is dual-containing, in which case
\begin{equation}\label{Eq:SymCSS}
  \tbf{H}(\mc{C}) = \begin{pmatrix} H(C) & 0 \\ 0 & H(C) \end{pmatrix}
\end{equation}
is a stabilizer parity check matrix for an $[[n, 2k-n, d]]$ CSS code $\mc{C}$.  We call CSS codes constructed in this manner \emph{symmetric} CSS codes; otherwise they are called \emph{asymmetric}.

%\IEEEpubidadjcol

%%%%%%%%%%%%%%%%%%%%%%%%%%%%%%%%%%%%%%%%%%%%%%%%%%%%%%%%%%%
%%%%%%%%%%%%%%%%%%%%%%%%%%%%%%%%%%%%%%%%%%%%%%%%%%%%%%%%%%%
\section{Finite Projective Planes and Regular Hyperovals}\label{PG(2,q)}

%%%%%%%%%%%%%%%%%%%%%%%%%%%%%%%%%%%%%%
%%%%%%%%%% Hyperoval Table %%%%%%%%%%%
%%%%%%%%%%%%%%%%%%%%%%%%%%%%%%%%%%%%%%
\begin{table*}
\centering
\begin{tabular}{|c|c|c|c|c|}\hline
 \                  & \                           & \                         & \# of secant lines & \# of skew lines\\
\# of points & \# of lines & \# of lines & intersecting a non- & intersecting a non-\\
 in $H_\mc{C}$ & secant to $H_\mc{C}$ & skew to $H_\mc{C}$ & hyperoval point & hyperoval point\\ \hline
$q+2$ & $\frac{q^2+3q+2}{2}$ & $\frac{q^2-q}{2}$ & $\frac{q+2}{2}$ & $\frac{q}{2}$ \\ \hline
\end{tabular}
\caption{Number of points and lines with respect to the hyperoval $H_{\mc{C}}$ and non-hyperoval points in $PG(2,q)$.}\label{Tab:HyperovalNums}
\end{table*}
%%%%%%%%%%%%%%%%%%%%%%%%%%%%%%%%%%%%%%
%%%%%%%%%% Hyperoval Table %%%%%%%%%%%
%%%%%%%%%%%%%%%%%%%%%%%%%%%%%%%%%%%%%%

\begin{figure}
  \centering
\includegraphics[width=0.15\textwidth]{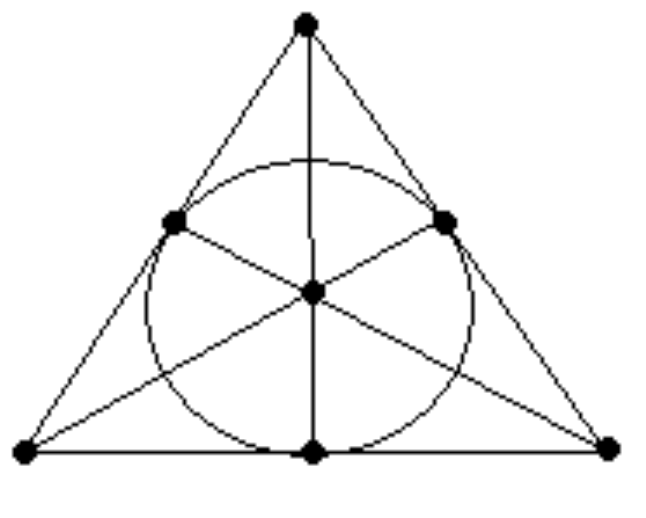}
\caption{The Fano Plane, $PG(2,2)$, is the simplest example of a finite projective plane.}
\label{Fig:fano}
\end{figure}

A \emph{finite projective plane} $PG(2,q)$ is a finite collection of points, along with subsets of points (lines) satisfying:
\begin{enumerate}
  \item Any two distinct points determine a unique line.
  
  \item Any two distinct lines determine a unique point.
  
  \item There exist four points, no three of which are colinear.
\end{enumerate}
Note that the second axiom implies that there are no parallel lines in this geometry.  See Figure \ref{Fig:fano} for an example of a finite projective plane.

The value $q$ is called the \emph{order} of the projective plane, and the following properties of $PG(2,q)$ can be determined \cite{PJCam}:
\begin{enumerate}
  \item Every line contains $q+1$ distinct points.
  
  \item Every point is incident with $q+1$ distinct lines.
  
  \item There are exactly $q^2+q+1$ points and $q^2+q+1$ lines in the plane.
\end{enumerate}

When $q$ is a prime power, then points and lines of $PG(2,q)$ can be represented as 1- and 2-dimensional subspaces, respectively, of the 3-dimensional vector space $V_3(q)$ over $\mbb{F}_q$.  Points are equivalence classes of the form $[x,y,z] \equiv \{(cx, cy, cz) \ | \ c \in \mbb{F}_q - \{0\} \}$, for $x$, $y$, and $z$ in $\mbb{F}_q$. While lines are 2-dimensional subspaces, each can be uniquely represented by its dual in the 1-dimensional subspace. Thus, to distinguish lines from points, we label points in brackets, e.g. $[x,y,z]$, and lines in perentheses, e.g. $(a,b,c)$.

%%%%%%%%%%%%%%%%%%%%%%%%%%%%%%%%%%%%%%%%%%%%%%%%%%%%%%%%%%%
\subsection{Conics, Hyperovals, and Subsets of lines}
A \emph{conic} $\mc{C}$ is a set of $q+1$ points in $PG(2,q)$ whose coordinates satisfy a non-degenerate quadradic equation, that is,
\begin{equation}
  \mc{C} := \{[x,y,z] : ax^2 + by^2 + cz^2 + fyz + gzx + hxy = 0 \},
\end{equation}
for some $a, b, c, f, g, h, \in \mbb{F}_q$. Conics have the property that no three-point subsets are colinear. As such, every line must be either \emph{skew} (i.e. intersects $\mc{C}$ at no points), \emph{tangent} (i.e. intersects $\mc{C}$ at one point), or \emph{secant} (i.e. intersects $\mc{C}$ at exactly two points). A well-known result in projective geometry is that when $q = 2^s$, all of the tangent lines are concurrent at a point outside the conic, called the \emph{nucleus}. If the nucleus is added to the conic, we obtain a \emph{regular hyperoval}, $H_\mc{C}$ (hereafter called simply a hyperoval). Lines tangent to the conic are then secant to the hyperoval, and hence all lines are either secant or skew to $H_\mc{C}$. Table \ref{Tab:HyperovalNums} lists the number of points and lines with respect to a hyperoval.

%%%%%%%%%%%%%%%%%%%%%%%%%%%%%%%%%%%%%%%%%%%%%%%%%%%%%%%%%%%
\subsection{Incidence Stuctures and Parity Checks}\label{Subsec:Incidence}
The \emph{incidence matrix} $M_\pi$ for $\pi = PG(2,q)$ is constructed by letting the columns correspond to points, and rows correspond to lines, with $M_{\pi_{i,j}} = 1$ when line $i$ contains point $j$, and $M_{\pi_{i,j}} = 0$ otherwise.  This matrix is sparse by construction. It was shown in \cite{KJCSmith69} that the rank of this matrix is $\begin{pmatrix}p+1\\2\end{pmatrix}^s + 1$, where $q = p^s$. In particular, when $p = 2$, this reduces to $3^s+1$.

The incidence matrix $M_\pi$ is well-suited to act as a parity check matrix for a classical LDPC code, as it is sparse, and any two rows have exactly one ``1'' in common position.  However, since this matrix is not self-dual, it must be adapted by adding a column of all ones, called the \emph{unit vector}, or $u$-vector, to it, denoting the new matrix by $M_\pi^\p$. Note that since every row of $M_\pi$ has odd weight, adding this vector does not affect the rank. Since $M_\pi^\p$ is self-dual, sparse, and any two rows have exactly two ``1''s in common position, it can be used in the construction of a parity check matrix for a quantum LDPC code.

We can likewise consider classical LDPC parity check matrices constructed by the incidence structures of subsets of points and lines in $PG(2,q)$. Such classical codes have been recently studied \cite{DrMel06, CHMel10}, but again, these incidence matrices must be adapted to make them self-orthogonal in order to use them to construct parity checks for QLDPC codes.

%%%%%%%%%%%%%%%%%%%%%%%%%%%%%%%%%%%%%%%%%%%%%%%%%%%%%%%%%%%
%%%%%%%%%%%%%%%%%%%%%%%%%%%%%%%%%%%%%%%%%%%%%%%%%%%%%%%%%%%
\section{Classical Self-Orthogonal LDPC Codes}\label{ClassicalLDPC}

Here we construct self-orthogonal parity check matrices for classical LDPC codes constructed from the incidence matrices from point-line subsets of $PG(2,q)$. In particular, we assume that $q=2^s$ so that we can use the subset structure based on hyperovals. We then study the properties of the corresponding classical codes. The results of this section are summarized in Table \ref{Tab:ClassicalResults}.

%%%%%%%%%%%%%%%%%%%%%%%%%%%%%%%%%%%%%%%%%%%%%%%%%%%%%%%%%%%
\subsection{All Points and All Lines, $M_\pi^\p$}\label{Subsec:M_pi^p}
We again let $M_\pi$ be the incidence matrix for $\pi = PG(2,q)$ for $q= 2^s$, and let $M_\pi^\p$ be the concatenation of $M_\pi$ with the $u$-vector, i.e. $M_\pi^\p = [M_\pi|\mathbf{1}]$. This matrix has $4^s+2^s+2$ columns and $4^s+2^s+1$ rows. As discussed in Section \ref{Subsec:Incidence}, the rank of this matrix is $3^s+1$.
\begin{proposition}\label{Prop:M_pi^p}
The matrix $M_\pi^\p$ is a parity check for a classical $[4^s+2^s+2, \ 4^s-3^s+2^s+1, \ 2^s+2]$ LDPC code.
\begin{proof}
The length and dimension are obvious. To prove minimum distance, we have two cases:
\begin{itemize}
\item[$A)$:] The last bit of a minimum weight codeword is a 0. In this case, the codeword will have the same weight as a minimum weight codeword for a code with classical parity check matrix given by $M_\pi$, which is known to be $2^s+2$ \cite{KLF01}.

\item[$B)$:] The last bit $c_n$ of a minimum weight codeword $c$ is a 1. Each additional 1-bit in $c$ will cause $c$ to be orthogonal to $2^s + 1$ rows of $M_\pi^\p$. Thus, in addition to $c_n$, the codeword $c$ needs at least $m$ additional 1-bits, where $m$ is the smallest integer such that $m(2^s+1)\geq 4^s+2^s+1$, which is determined to be $2^s+1$. Thus, if $c_n = 1$, then the weight of $c$ is lower bounded by $2^s+2$, completing the proof.
\end{itemize}
\end{proof}
\end{proposition}

Note that, we can alternatively prove results on minimum weight from a graph-theoretic approach.  In particular, $M_\pi^\p$ can be graphically viewed as the incidence matrix for a projective plane $\pi = PG(2,q)$ with an additional point $u$ through which every line intersects.  Then a codeword is graphically viewed as a collection of points in this ``extended projective plane'' $\pi^\p$, through which every line intersects an even number of times, since this would correspond to a vector having an even number of 1-bits in common position with each row of the parity check, and hence orthogonal to each row.  In particular, if the codeword has a 0 at the $u$-column, then we reduce ourselves to studying a collection of points $\mc{S}$ in $\pi$ satisfying this. However, if the codeword has a 1 at the $u$-column, then the collection of points contains the point $u$ through which every line intersects.  Then the remaining points $\mc{S}$ are all in $\pi$ such that every line intersects $\mc{S}$ an odd number of times. The weight of the codeword is given by $|\mc{S}|$ or $|\mc{S}|+1$ depending on whether the codeword has a 0 or a 1 at the end, respectively. Thus, we can determine the minimum weight of a code by finding a minimal set of points $\mc{S}$ in $PG(2,q)$ in each case. Such an approach will be used for many of the proofs in the following constructions.
%%%%%%%%%%%%%%%%%%%%%%%%%%%%%%%%%%%%%%%%%%%%%%%%%%%%%%%%%%%
\subsection{Skew Lines and Non-Hyperoval Points, $H(C_{sk})$}\label{Subsec:H(C_sk)}

%%%%%%%%%%%%%%%%%%%%%%%%%%%%%%%%%%%%%%
%%%%%%%Classical Results Table%%%%%%%%
%%%%%%%%%%%%%%%%%%%%%%%%%%%%%%%%%%%%%%
\begin{table}[!t]
  \centering
\begin{tabular}{c||c|c|c|}
 \         & $n$         &      $k$           & $d$\\ \hline \hline
$C_\pi^\p$ & $4^s+2^s+2$ & $4^s-3^s+2^s+1$    & $2^s+2$\\ \hline 
 \         & \           & $4^s-3^s-1 \leq k$ & $2^{s-1}+1$  \\
$C_{sk}$   & $4^s$       & $\leq 4^s-3^s+2^s$ & $\leq d \leq 2^s$ \\ \hline
 \         &     \       &      \             & $2^{s-1}+2$ \\
$C_{seA}$  & $4^s+2^s+2$ & $4^s-3^s+2^s+1$    & $\leq d \leq 2^s+2$ \\ \hline
$C_{se}$   & $4^s$       & $4^s-3^s+2^s+1$    & $2^{s-1}+2 \leq d \leq 2^s+2$ \\ \hline
\end{tabular}
\caption{Summary of classical LDPC codes constructed from $PG(2,2^s)$ and subsets with respect to a hyperoval.}\label{Tab:ClassicalResults}
\end{table}
%%%%%%%%%%%%%%%%%%%%%%%%%%%%%%%%%%%%%%
%%%%%%%Classical Results Table%%%%%%%%
%%%%%%%%%%%%%%%%%%%%%%%%%%%%%%%%%%%%%%

Suppose we restrict ourselves to the incidence structure formed by only the lines skew to a hyperoval, along with the $u$-vector. In such a case, we can find a weight one codeword by letting the last bit be 0, and letting $\mc{S}$ consist of only a hyperoval point. Since skew lines intersect the hyperoval nowhere, $\mc{S}$ corresponds to a codeword. A code of minimum weight 1 is useless, so we will instead delete the columns corresponding to hyperoval points. Thus, we let $H(C_{sk})$ be the incidence matrix whose rows correspond to skew lines and whose columns correspond to non-hyperoval points in$PG(2,q)$, concatenated with the $u$-vector. This matrix will have $\frac{q^2-q}{2}$ rows and $q^2$ columns.

\begin{proposition}\label{Prop:H(C_sk)}
The matrix $H(C_{sk})$ is a parity check matrix for a classical $[4^s, k_{sk}, d_{sk}]$ LDPC code $C_{sk}$, where $4^s-3^s-1 \leq k_{sk} \leq 4^s - 3^s + 2^s$ and $2^{s-1}+1 \leq d_{sk} \leq 2^s$.
\begin{proof}
We know that $\dim(M_\pi^\p) = 3^s+1$. Since skew lines never intersect hyperoval points, removal of the columns corresponding to hyperoval points does not change the dimension. In \cite{CHMel10} it was shown that removing rows corresponding to lines secant to the conic will not affect the dimension. Removing the lines tangent to the conic (i.e. the remaining lines secant to the hyperoval) gives us the matrix $H(C_{sk})$, but may possibly decrease the dimension. Thus $3^s-2^s \leq \dim(H(C_{sk})) \leq 3^s+1$, from which we determine the bounds on the code dimension.

To prove minimum distance, we again consider two cases, namely, if the last bit of a codeword is a 0 or a 1.
\begin{itemize}
\item[$A)$:]  Suppose the codeword has a 0 at the $u$-column. Then we view the codeword as a collection of non-hyperoval points $\mc{S}$ in $PG(2,q)$ such that any skew line intersects $\mc{S}$ in an even number of places. Let $p$ be a point in $\mc{S}$. Then it must be intersected by $q/2$ skew lines. Then for each of these skew lines, $\mc{S}$ must have an additional point through which the line intersects. Thus, $|\mc{S}| \geq \frac{q}{2} + 1$.
\item[$B)$:]  Suppose the codeword has a 1 at the $u$-column. Then we view the codeword as a collection of non-hyperoval points $\mc{S}$ in $PG(2,q)$ such that any skew line intersects $\mc{S}$ in an odd number of places, namely, at least once. Since any two lines in $PG(2,q)$ intersect at exactly one point, we obtain a minimum when we choose $\mc{S}$ to be a line. Since hyperoval points are removed, secant lines have the fewest number of non-hyperoval points in $PG(2,q)$, namely $q-1$. Thus, if the codeword has a 1 in the $u$-column, then it must have weight at least $|\mc{S}|+1 = q$. Since $\mc{S}$ was found constructively, codewords of such weight do exist, and hence this acts as our upper bound on the minimum weight.
\end{itemize}
\end{proof}
\end{proposition}

%%%%%%%%%%%%%%%%%%%%%%%%%%%%%%%%%%%%%%%%%%%%%%%%%%%%%%%%%%%
\subsection{Secant Lines and All Points, $H(C_{seA})$}\label{Subsec:H(C_seA)}

We now consider classical LDPC codes whose incidence matrices are constructed by removing the rows in $M_\pi^\p$ corresponding to lines skew to a given hyperoval, leaving only the rows corresponding to the lines secant to the hyperoval. This matrix, denoted $H(C_{seA})$, will have $q^2+q+2$ columns and $(q^2+3q+2)/2$ rows.

\begin{proposition}\label{Prop:H(C_seA)}
The matrix $H(C_{seA})$ is a parity check matrix for a classical $[4^s+2^s+2, \ 4^s-3^s+2^s+1, \ d_{seA}]$ LDPC code $C_{seA}$, where $2^{s-1}+2 \leq d_{seA} \leq 2^s+2$.

\begin{proof}
We know $\dim(M_\pi^\p) = 3^s+1$. It is also known \cite{CHMel10} that removing the rows corresponding to skew lines (thereby giving us $H(C_{seA})$) does not change the rank of this matrix. Thus $\dim(H(C_{seA})) = 3^s+1$. Since the length $n$ is $4^s+2^s+2$, we solve for $\dim(C_{seA}) = n-\dim(H(C_{seA}))$ to obtain our result.

To prove the bounds on the minimum distance, we again have the following two cases:
\begin{itemize}
\item[$A)$:] Suppose the codeword has a 0 at the $u$-column. Then we view the codeword as a collection of points $\mc{S}$ in $PG(2,q)$ such that any secant line intersects $\mc{S}$ in an even number of places. Let $p$ be a point in $\mc{S}$. Then it must be intersected by $(q+2)/2$ secant lines. Then for each of these secant lines, $\mc{S}$ must have an additional point through which the line intersects. Thus, $|\mc{S}| \geq \frac{q+2}{2} + 1 = \frac{q}{2} + 2 = 2^{s-1} + 2$.
\item[$B)$:]  Suppose the codeword has a 1 at the $u$-column. Then it corresponds to a collection of points $\mc{S}$ in $PG(2,q)$ such that every line secant to the hyperoval intersects it an odd number of times, namely, at least once. Since each line in $PG(2,q)$ intersects all other lines exactly once, we obtain a minimum when we choose the points in $\mc{S}$ to be a collection of $q+1$ points that make up a line. Thus, if a codeword has a 1 at the $u$-column, then it must have weight at least $|\mc{S}|+1 = q+2 = 2^s+2$. Since $\mc{S}$ was found constructively, codewords of such weight do exist, and hence this acts as the upper bound on the minimum weight.
\end{itemize}
\end{proof}
\end{proposition}

%%%%%%%%%%%%%%%%%%%%%%%%%%%%%%%%%%%%%%%%%%%%%%%%%%%%%%%%%%%
\subsection{Secant Lines and Non-Hyperoval Points, $H(C_{se})$}\label{Subsec:H(C_se)}

While removing hyperoval points was not necessary to obtain good classical codes constructed from secant lines, we nevertheless remove them here for reasons that will become apparent later.  Let $H(C_{se})$ be the matrix formed by removing from $M_\pi^\p$ the columns corresponding to hyperoval points and the rows corresponding to lines skew to the hyperoval.  This matrix will have $(q^2+3q+2)/2$ rows and $q^2$ columns.

\begin{proposition}\label{Prop:H(C_se)}
The matrix $H(C_{se})$ is a parity check matrix for a classical $[4^s, \ 4^s-3^s+2^s+1, \ d_{se}]$ LDPC code $C_{se}$, where $2^{s-1}+2 \leq d_{se} \leq 2^s+2$.
\begin{proof}
We know $\dim(M_\pi) = 3^s+1$. In \cite{CHMel10} it was shown that columns corresponding to hyperoval points and rows corresponding to skew lines can be removed without affecting the dimension. Since the weight of each row of the resulting matrix is odd, we can add the $u$-vector, to obtain the matrix $H(C_{se})$, without affecting the rank. Thus, $\dim(H(C_{se})) = 3^s+1$, from which we determine the dimension of $C_{se}$.

To prove the minimum distance, we have the following two cases:
\begin{itemize}
  \item[$A)$:]  Suppose a codeword has a 0 at the $u$-column. Then it corresponds to a collection of non-hyperoval points $\mc{S}$ in $PG(2,q)$ such that every secant line intersects $\mc{S}$ in an even number of points. Let $p \in \mc{S}$. Since it is intersected by $(q+2)/2$ secant lines, we must have at least one additional point in $\mc{S}$ for each line to intersect. Thus $|\mc{S}| \geq \frac{q}{2}+2 = 2^{s-1}+2$.
  \item[$B)$:]  Suppose a codeword has a 1 at the $u$-column. The proof continues in a similar fashion to that of Proposition \ref{Prop:H(C_seA)}. A set of points $\mc{S}$ in $PG(2,q)$ corresponding to a codeword with a $1$ at the $u$-column will be minimal when it intersects every secant line exactly once in non-hyperoval points, which occurs when $\mc{S}$ is a collection of $q+1$ points corresponding to a skew line. Thus, $|\mc{S}|+1 = 2^s+2$. Again, since such codewords do exist, this weight acts as an upper bound on the minimum weight of the code.
\end{itemize}
\end{proof}
\end{proposition}

%%%%%%%%%%%%%%%%%%%%%%%%%%%%%%%%%%%%%%%%%%%%%%%%%%%%%%%%%%%
%%%%%%%%%%%%%%%%%%%%%%%%%%%%%%%%%%%%%%%%%%%%%%%%%%%%%%%%%%%
\section{Quantum LDPC Codes}\label{QLDPC}

Using the results from Section \ref{ClassicalLDPC}, we construct parity check matrices for QLDPC codes using the asymmetric and symmetric CSS constructions \eqref{Eq:AsymCSS} and \eqref{Eq:SymCSS}.

%%%%%%%%%%%%%%%%%%%%%%%%%%%%%%%%%%%%%%%%%%%%%%%%%%%%%%%%%%%
\subsection{Symmectric QLDPC Codes from All Points and All Lines}

The symmetric QLDPC codes $\mc{C}_\pi$ constructed here have as a parity check matrix $H(\mc{C}_\pi)$ of the form
\begin{equation}\label{Eq:H(C_pi)}
  H(\mc{C}_\pi) = \begin{bmatrix} M_\pi^\p & 0\\ 0 & M_\pi^\p \end{bmatrix},
\end{equation}
where $M_\pi^\p$ is as defined in Section \ref{Subsec:M_pi^p}. Recall that $M_\pi^\p$ is self-orthogonal by construction, and is a parity check matrix for a classical LDPC code, and hence $H(\mc{C}_\pi)$ is well-defined.

\begin{theorem}\label{Thm:H(C_pi)}
Given a finite projective plane $\pi = PG(2,2^s)$ for some positive integer $s$, the matrix $H(\mc{C}_\pi)$ in \eqref{Eq:H(C_pi)} is a parity check matrix for an $[[n, \ 2k-n, \ D]]$ QLDPC code $\mc{C}_\pi$, where $n = 4^s+2^s+2$, $k = 4^s-3^s+2^s+1$, and $D = 2^s+2$.
\begin{proof}
This follows immediately from Proposition \ref{Prop:M_pi^p} and the properties of symmetric CSS codes.
\end{proof}
\end{theorem}

Observe that this code has a rate $\frac{2k-n}{n}$ that rapidly increases with the length, and a minimum distance that increases on the order $\sqrt{n}$. The number of stabilizers used for error correction, given by the number of rows in the parity check, is almost $2n$.

%%%%%%%%%%%%%%%%%%%%%%%%%%%%%%%%%%%%%%%%%%%%%%%%%%%%%%%%%%%
\subsection{Asymmetric QLDPC Codes}\label{Subsec:asym}

The asymmetric QLDPC codes $\mc{C}_{asym}$ constructed here have as a parity check matrix $H(\mc{C}_{asym})$ of the form
\begin{equation}\label{Eq:H(C_asym)}
  H(\mc{C}_{asym}) = \begin{bmatrix} H(C_{sk}) & 0\\ 0 & H(C_{se}) \end{bmatrix},
\end{equation}
where $H(C_{sk})$ and $H(C_{se})$ are as defined in Sections \ref{Subsec:H(C_sk)} and \ref{Subsec:H(C_se)}, respectively. Note that $H(C_{sk})$ and $H(C_{se})$ are orthogonal by construction, and both have the same block length. Moreover, each is a parity check for a classical LDPC code, making $H(\mc{C}_{asym})$ a well-defined parity check for a QLDPC code.

\begin{theorem}\label{Thm:C_asym}
Given a finite projective plane $PG(2,2^s)$ for some positive integer $s$, the matrix $H(\mc{C}_{asym})$ in \eqref{Eq:H(C_asym)} is a parity check matrix for a $[[4^s, K, D]]$ QLDPC code $\mc{C}_{asym}$, where $4^s-2\cdot 3^s+2 \leq K \leq 4^s-2\cdot 3^s + 2^s-1$, and $D \geq 2^{s-1}+1$.
\begin{proof}
The length is determined by observing that $H(C_{sk})$ and $H(C_{se})$ are parity checks for classical codes of length $4^s$. We obtain the dimension $K$ and minimum distance $D$ from Propositions \ref{Prop:H(C_sk)} and \ref{Prop:H(C_se)}, and the fact that $K = \dim(\mc{C}_{asym}) = k_{sk} + k_{se} - n$, and minimum distance $D$ is bounded below by $\min(d_{sk}, d_{se})$, where $k_{sk}$, $k_{se}$, $d_{sk}$ and $d_{se}$ are respectively the dimensions and minimum distances of the classical codes $C_{sk}$ and $C_{se}$, respectively.
\end{proof}
\end{theorem}

While the bound on the dimension may be coarse, it is important to observe that the rate of these codes nevertheless increases rapidly with the code length $n$. The code will have only $n + \sqrt{n} + 1$ parity checks, and very few four-cycles.

%%%%%%%%%%%%%%%%%%%%%%%%%%%%%%%%%%%%%%%%%%%%%%%%%%%%%%%%%%%
\subsection{Symmectric QLDPC Codes from Skew Lines}

The parity check matrix $H(\mc{C}_{symSK})$ for the symmetric QLDPC codes $\mc{C}_{symSK}$ constructed here have the form
\begin{equation}\label{Eq:H(C_symSK)}
  H(\mc{C}_{symSK}) = \begin{bmatrix} H(C_{sk}) & 0\\ 0 & H(C_{sk}) \end{bmatrix},
\end{equation}
where $H(C_{sk})$ is the self-orthogonal classical LDPC parity check matrix defined in Section \ref{Subsec:H(C_sk)}.

\begin{theorem}\label{Thm:C_symSK)}
Given a finite projective plane $PG(2,2^s)$ for some positive integer $s$, the matrix $H(\mc{C}_{symSK})$ defined in \eqref{Eq:H(C_symSK)} is a parity check matrix for a $[[4^s, \ 2k_{sk}-4^s, \ D]]$ QLDPC code $\mc{C}_{symSK}$, where $4^s-3^s-1 \leq k_{sk} \leq 4^s - 3^s + 2^s$, and $D \geq 2^{s-1}+1$.
\begin{proof}
This follows immediately from Proposition \ref{Prop:H(C_sk)} and the properties of symmetric CSS codes.
\end{proof}
\end{theorem}

Although the bound on the dimension of these codes is not as tight as that of the asymmetric codes described in Section \ref{Subsec:asym}, these codes nevertheless have a fast rate that increases with the length $n$.  The bound on the minimum distance is the same as for the asymmetric codes, showing that these, too, describe fast-rate QLDPC codes with good minimum distance.  The codes will have $n - \sqrt{n}$ parity checks and few four-cycles.

%%%%%%%%%%%%%%%%%%%%%%%%%%%%%%
%%%%%%%%%%%%%%%%%%%%%%%%%%%%%%
%%%%Table of QLDPC Results%%%%
%%%Placed here for better--%%%
%%%% output visualization%%%%%
%%%%%%%%%%%%%%%%%%%%%%%%%%%%%%
%%%%%%%%%%%%%%%%%%%%%%%%%%%%%%
\begin{table*}[!t]
\centering
  \begin{tabular}{|c|c|c|c|c|c|} \hline
   \textbf{Code} & \textbf{CSS Type} & \textbf{Length} & \textbf{Dimension}            & \textbf{Minimum}      & \textbf{Number of} \\
    \            & \                 & \               & \                             & \textbf{Distance}     & \textbf{Stabilizers} \\
    \            & \                 & \               & \                             & \textbf{(Lower Bound)}& \                    \\ \hline \hline
   $C_\pi$       & Symmetric         & $4^s + 2^s + 2$ & $4^s - 2\cdot 3^s + 2^s$      & $2^s + 2$             & $2^{2s + 1} + 2^{s+1} + 2$\\ \hline
   $C_{asym}$    & Asymmetric        & $4^s$           & $4^s - 2\cdot 3^s + 2 \leq K$ & $2^{s-1} + 1$         & $4^s + 2^s + 1$\\
    \            & \                 &     \           & $\leq 4^s - 2\cdot 3^s+2^s-1$ & \                     &  \          \\ \hline
   $C_{symSK}$   & Symmetric         & $4^s$           & $4^s-2\cdot 3^s- 2 \leq K$    & $2^{s-1}+1$           & $4^s - 2^s$ \\
    \            & \                 & \               & $\leq 4^s-2\cdot 3^s+2^{s+1}$ & \                     & \           \\ \hline
   $C_{symSE}$   & Symmetric         & $4^s + 2^s + 2$ & $4^s - 2\cdot 3^s + 2^s$      & $2^{s-1}+2$           & $4^s+3\cdot 2^s+2$ \\ \hline
  \end{tabular}
  \caption{QLDPC Code Parameters for Parity Checks Constructed from $PG(2,2^s)$}\label{Tab:summary}
\end{table*}
%%%%%%%%%%%%%%%%%%%%%%%%%%%%%%
%%%%%%%%%%%%%%%%%%%%%%%%%%%%%%
%%%%Table of QLDPC Results%%%%
%%%Placed here for better--%%%
%%%% output visualization%%%%%
%%%%%%%%%%%%%%%%%%%%%%%%%%%%%%
%%%%%%%%%%%%%%%%%%%%%%%%%%%%%%

%%%%%%%%%%%%%%%%%%%%%%%%%%%%%%%%%%%%%%%%%%%%%%%%%%%%%%%%%%%
\subsection{Symmetric QLDPC Codes from Secant Lines}

Two different classical LDPC codes constructed from secant lines were discussed in Section \ref{ClassicalLDPC}, namely those whose parity checks $H(C_{seA})$ were constructed from all points, and those whose parity checks $H(C_{se})$ had the columns corresponding to hyperoval points removed.  Although $H(C_{se})$ is orthogonal to $H(C_{sk})$, it is not self-orthogonal, while $H(C_{seA})$ is. Thus, the parity check matrix $H(\mc{C}_{symSE})$ for the symmetric QLDPC codes $\mc{C}_{symSE}$ constructed here have the form
\begin{equation}\label{Eq:H(C_symSE)}
  H(\mc{C}_{symSE}) = \begin{bmatrix} H(C_{seA}) & 0\\ 0 & H(C_{seA}) \end{bmatrix}.
\end{equation}

\begin{theorem}\label{Thm:H(C_symSE)}
Given a finite projective plane $PG(2,2^s)$, for some positive integer $s$, the matrix $H(\mc{C}_{symSE})$ in \eqref{Eq:H(C_symSE)} is a parity check matrix for an $[[n, 2k_{seA}-n, D]]$ QLDPC code $\mc{C}_{symSE}$, where $n = 4^s+2^s+2$, $k_{seA} = 4^s-3^s+2^s+1$, and $D \geq 2^{s-1}+2$.
\begin{proof}
Here $k_{seA} = \dim(C_{seA})$. The rest follows from Proposition \ref{Prop:H(C_seA)} and the properties of symmetric CSS codes.
\end{proof}
\end{theorem}

Note that the length and dimension (and therefore the rate) of this code is the same as that of $\mc{C}_\pi$, while the minimum distance and number of stabilizers is roughly half that of $\mc{C}_\pi$.

\section{Conclusion}
Table \ref{Tab:summary} gives a summary of the results for each of the QLDPC codes discussed in this report.

While many of the parameters are not exact for many of these codes, the bounds nevertheless indicate that each of these codes are at least comparable to most quantum LDPC codes in the literature. In fact, as previously mentioned, many of these parameters are completely unknown for the other quantum LDPC codes. 

While further research is necessary to determine exact values of minimum distance and dimension of most of these codes, it is nevertheless established that the projective plane is a very useful tool in the construction of quantum low-density parity check codes.  Similar techniques can also be used to construct QLDPC codes from $PG(m, p^s)$ for $m > 2$ and/or $p$ an odd prime.  Note that in the case of $PG(2,p^s)$ where $p$ is an odd prime, the $u$-column is not necessarily linearly dependent on the columns of the corresponding incidence matrix, making it much more difficult to determine dimensions of corresponding QLDPC codes.  However, this is resolved if in addition to concatenating the $u$-column to the incidence matrix of the projective plane, you also concatenate an identity matrix to it.  This would change the parameters in a known manner. Additionally, if $p$ is odd, then the point-line subsets should be taken with respect to a conic, as we lose the regular hyperoval structure present when $p$ is even.

% or


%\appendices

%\section{}

\section*{Acknowledgment}

The author would like to thank James Troupe, Keye Martin, Keith Mellinger, and Geir Agnarsson for useful discussions.

\ifCLASSOPTIONcaptionsoff
  \newpage
\fi



\bibliographystyle{IEEEtran}

\bibliography{bibfile}

\begin{thebibliography}{10}
\providecommand{\url}[1]{#1}
\csname url@samestyle\endcsname
\providecommand{\newblock}{\relax}
\providecommand{\bibinfo}[2]{#2}
\providecommand{\BIBentrySTDinterwordspacing}{\spaceskip=0pt\relax}
\providecommand{\BIBentryALTinterwordstretchfactor}{4}
\providecommand{\BIBentryALTinterwordspacing}{\spaceskip=\fontdimen2\font plus
\BIBentryALTinterwordstretchfactor\fontdimen3\font minus
  \fontdimen4\font\relax}
\providecommand{\BIBforeignlanguage}[2]{{%
\expandafter\ifx\csname l@#1\endcsname\relax
\typeout{** WARNING: IEEEtran.bst: No hyphenation pattern has been}%
\typeout{** loaded for the language `#1'. Using the pattern for}%
\typeout{** the default language instead.}%
\else
\language=\csname l@#1\endcsname
\fi
#2}}
\providecommand{\BIBdecl}{\relax}
\BIBdecl

\bibitem{JFarin12}
J.~M. Farinholt, ``Classes of high-performance quantum ldpc codes from finite
  projective geometries,'' Master's Thesis, George Mason University, 2012.

\bibitem{Gal62}
R.~G. Gallager, ``Low density parity check codes,'' \emph{Transactions of the
  IRE Professional Group on Information Theory}, vol. IT-8, pp. 21--28, January
  1962.

\bibitem{KLF01}
Y.~Kou, S.~Lin, and M.~P.~C. Fossorier, ``Low density parity check codes based
  on finite geometries: A rediscovery and new results,'' \emph{IEEE Trans.
  Inform. Theory}, vol.~47, pp. 2711--2736, 2001.

\bibitem{DrMel06}
Droms, Meyer, and K.~E. Mellinger, ``Ldpc codes generated by conics in the
  classical projective plane,'' \emph{Designs, Codes, and Cryptography},
  vol.~40, no.~3, 2006.

\bibitem{CHMel10}
Castleberry, K.~Hunsberger, and K.~E. Mellinger, ``Ldpc codes arising from
  hyperovals,'' \emph{Bull. Inst. Combin. Appl.}, vol.~58, pp. 59 -- 72, 2010.

\bibitem{Shan48}
C.~E. Shannon, ``A mathematical theory of communication,'' \emph{Bell System
  Technical Journal}, vol.~27, pp. 379 -- 423, 623 -- 656, 1948.

\bibitem{Shor95}
P.~W. Shor, ``Scheme for reducing decoherence in quantum computer memory,''
  \emph{Phys. Rev. A}, vol.~52, pp. 2493--2496, 1995.

\bibitem{CalShor96}
A.~R. Calderbank and P.~W. Shor, ``Good quantum error-correcting codes exist,''
  \emph{arXiv:quant-ph/9512032}, 1996.

\bibitem{Steane}
A.~M. Steane, ``Multiple particle interference and quantum error correction,''
  \emph{Proc. R. Soc. Lond. A}, vol. 452, no. 2551, 1996.

\bibitem{Stea96}
------, ``Error correcting codes in quantum theory,'' \emph{Phys. Rev. Lett.},
  vol.~77, pp. 793 -- 797, 1996.

\bibitem{Gott96}
D.~Gottesman, ``Class of quantum error-correcting codes saturating the quantum
  hamming bound,'' \emph{Phys. Rev. A}, vol.~54, pp. 1862--1868, 1996.

\bibitem{CalRainsShorSloane}
A.~R. Calderbank, E.~M. Rains, P.~W. Shor, and N.~J.~A. Sloane, ``Quantum error
  correction and orthogonal geometry,'' \emph{Phys. Rev. Lett.}, vol.~78,
  no.~3, 1997.

\bibitem{Pos01}
Postol, ``A proposed quantum low density parity check code,''
  \emph{arXiv:quant-ph/0108131}, 2001.

\bibitem{MkMf04}
MacKay, Mitchison, and McFadden, ``Sparse-graph codes for quantum error
  correction,'' \emph{IEEE Transactions on Information Theory}, vol.~50,
  no.~10, 2004.

\bibitem{TilZem09}
\BIBentryALTinterwordspacing
J.-P. Tillich and G.~Z{\'e}mor, ``Quantum ldpc codes with positive rate and
  minimum distance proportional to n 1/2,'' in \emph{Proceedings of the 2009
  IEEE international conference on Symposium on Information Theory - Volume 2},
  ser. ISIT'09.\hskip 1em plus 0.5em minus 0.4em\relax Piscataway, NJ, USA:
  IEEE Press, 2009, pp. 799--803. [Online]. Available:
  \url{http://dl.acm.org/citation.cfm?id=1701275.1701299}
\BIBentrySTDinterwordspacing

\bibitem{Aly09}
S.~Aly, ``A class of quantum ldpc codes constructed from finite geometries,''
  in \emph{Proc. 2008 IEEE Global Communication}, ser. Globecom '08, New
  Orleans, LA, USA, 2008.

\bibitem{PJCam}
P.~J. Cameron, \emph{Combinatorics: Topics, Techniques, Algorithms}.\hskip 1em
  plus 0.5em minus 0.4em\relax Cambridge University Press, 1994.

\bibitem{KJCSmith69}
K.~J.~C. Smith, ``On the $p$-rank if the incidence matrix of points and
  hyperplanes in a finite projective geometry,'' \emph{J. Comb. Theory},
  vol.~7, pp. 122 -- 129, 1969.

\end{thebibliography}

%\begin{biography}[{\includegraphics[width=1in,height=1.25in,clip,keepaspectratio]{mshell}}]{Michael Shell}

%\end{IEEEbiography}


%\newpage



%\vfill


\end{document}